\documentclass{PoS}

\usepackage{graphicx}

\newcommand{\note}[1]{}                    

\newcommand{\mnote}[1]{}                   

\newcommand{\sixth}{\mbox{\small $\frac{1}{6}$}}         
\newcommand{\half}{\mbox{\small $\frac{1}{2}$}}          
\newcommand{\quart}{\mbox{\small $\frac{1}{4}$}}         
\newcommand{\lat}{\mbox{\tiny $lat$}}                    

\def\lsim{\mathrel{\rlap{\lower4pt\hbox{\hskip1pt$\sim$}}
    \raise1pt\hbox{$<$}}}                
\def\gsim{\mathrel{\rlap{\lower4pt\hbox{\hskip1pt$\sim$}}
    \raise1pt\hbox{$>$}}}                

\title{
\vspace*{-1.25cm}
\begin{minipage}{\textwidth}
\begin{flushright}
\texttt{\footnotesize
PoS(LATTICE2015)264   \\
ADP-15-51/T953        \\
DESY 15-224           \\
Edinburgh 2015/32     \\
Liverpool LTH 1071    \\
}
\end{flushright}
\end{minipage}\\[15pt]
\vspace*{+1.25cm}
       Determining the scale in Lattice QCD}

\ShortTitle{Wilson flow scales}

\author{V.~G. Bornyakov$^{\,a}$,
        \speaker{R. Horsley}$^{\,b}$,
        R. Hudspith$^c$,
        Y. Nakamura$^d$,
        H. Perlt$^e$,
        D. Pleiter$^{f}$, 
        P.~E.~L. Rakow$^g$,
        G. Schierholz$^h$,
        A. Schiller$^e$,
        H. St\"uben$^i$
        and J.~M. Zanotti$^j$ \\
        \llap{$^a$} Institute for High Energy Physics, 
                    Protvino,
                    142281 Protvino, Russia, \\
        \llap{\phantom{$^a$}} 
                    Institute of Theoretical and Experimental Physics,
                    Moscow,
                    117259 Moscow, Russia, \\
        \llap{\phantom{$^a$}} 
                    School of Biomedicine, 
                    Far Eastern Federal University,
                    690950 Vladivostok, Russia \\
        \llap{$^b$} School of Physics and Astronomy,
                    University of Edinburgh,
                    Edinburgh  EH9 3FD, UK \\
        \llap{$^c$} Department of Physics and Astronomy,
                    York University,
                    Toronto, ON Canada M3J 1P3 \\
        \llap{$^d$} RIKEN Advanced Institute for Computational Science,
                    Kobe, Hyogo 650-0047, Japan \\
        \llap{$^e$} Institut f\"ur Theoretische Physik,
                    Universit\"at Leipzig, 04109 Leipzig, Germany \\
        \llap{$^f$} JSC, Forschungszentrum J\"ulich,
                    52425 J\"ulich, Germany \\
        \llap{\phantom{$^f$}} 
                    Institut f\"ur Theoretische Physik,
                    Universit\"at Regensburg, 93040 Regensburg, Germany \\
        \llap{$^g$} Theoretical Physics Division,
                    Department of Mathematical Sciences,
                    University of Liverpool, \\
        \llap{\phantom{$^g$}} 
                    Liverpool L69 3BX, UK \\
        \llap{$^h$} Deutsches Elektronen-Synchrotron DESY,
                    22603 Hamburg, Germany \\
        \llap{$^i$}  Universit\"at Hamburg, Regionales Rechenzentrum,
                    20146 Hamburg, Germany \\
        \llap{$^j$} CSSM, Department of Physics,
                    University of Adelaide, Adelaide SA 5005, Australia \\
        E-mail: \email{rhorsley@ph.ed.ac.uk} }

\author{QCDSF-UKQCD Collaborations}

\abstract{We discuss scale setting in the context of 2+1 dynamical
          fermion simulations where we approach the physical point
          in the quark mass plane keeping the average quark mass constant.
          We have simulations at four beta values, and after determining
          the paths and lattice spacings, we give an estimation of the
          phenomenological values of various Wilson flow scales.}

\FullConference{The 33rd International Symposium on Lattice Field Theory\\
		14 -18 July 2015\\
		Kobe International Conference Center, Kobe, Japan}

\begin{document}


\section{Singlet quantities}
\label{singlet}


Numerical lattice QCD simulations determine mass (or other) ratios
but not the scale itself, which has to be determined from experiment.
A hadron mass such as the proton mass or decay constant such as
the pion decay constant are often used for this purpose.
We discuss here the advantages of setting the scale using
a flavour-singlet quantity, which in conjunction with simulations
keeping the average quark mass constant allow $SU(3)$ flavour breaking
expansions to be used. This is illustrated using $2+1$ flavour clover
fermions, and in addition a determination of the Wilson flow scales,
$\sqrt{t_0^{\exp}}$ and $w_0^{\exp}$ is given.

This talk is based on \cite{bornyakov15a}, where further details can
be found.

Dynamical simulations start with some values of the quark masses
and then extrapolate along some path in $(u, d, s)$ space%
\footnote{Practically we consider mass degenerate $u$ and $d$ quarks,
when $m_u = m_d \equiv m_l$ but the discussion here is more general.}
to the physical point. The strategy we have adopted here,
\cite{bietenholz10a,bietenholz11a} is to start at a point on the
$SU(3)$ flavour symmetric line, when all the quark masses are equal
\begin{eqnarray}
   (m_0, m_0, m_0) \to (m_u^*, m_d^*, m_s^*) \,,
\end{eqnarray}
and to keep the singlet quark mass $\overline{m}$ constant
\begin{eqnarray}
   \overline{m}
      = {1 \over 3}\,\left( m_u + m_d + m_s \right)
      = \mbox{const.} \equiv m_0\,.
\end{eqnarray}
This allows an $SU(3)_F$ flavour symmetry breaking expansion
for masses and matrix elements. The expansion parameter is naturally
the distance from the $SU(3)$ flavour plane, parametrised by
\begin{eqnarray}
   \delta m_q  = m_q - \overline{m} \,.
\end{eqnarray}
This has the trivial constraint
\begin{eqnarray}
   \delta m_u+\delta m_d+\delta m_s = 0 \,.
\end{eqnarray}
The expansion coefficients are functions of $\overline{m}$ only
so provided $\overline{m}$ is kept constant they remain unaltered
whether we have mass degenerate $u$ and $d$ quarks or not.
This opens the possibility of determining isospin breaking
quantities from just $2+1$ simulations. The plane (or path) is
called `unitary' if we expand in both the same sea and valence quarks masses.

Consider now a flavour singlet quantity $X_S(m_u, m_d, m_s)$
which by definition is invariant under $u$, $d$, $s$ permutations.
This has a stationary point about the $SU(3)$ flavour symmetric line.
For upon expanding a flavour singlet quantity about a point on the
$SU(3)$-flavour line we have
\begin{eqnarray}
   \lefteqn{
      X_S( \overline{m} + \delta m_u,
           \overline{m} + \delta m_d,
           \overline{m} + \delta m_s ) }
     & &                                                     \nonumber \\
     &=& X_S(\overline{m}, \overline{m}, \overline{m})
      + \left. { \partial X_S \over \partial m_u } \right|_0 \,
                                         \delta m_u
      + \left. { \partial X_S \over \partial m_d } \right|_0 \,
                                         \delta m_d
      + \left. { \partial X_S \over \partial m_s } \right|_0 \,
                                         \delta m_s
      + O( (\delta m_q)^2 ) \,.
\label{stationary}
\end{eqnarray}
However on this line all the above derivatives are 
equal and thus we have
\begin{eqnarray}
   X_S( \overline{m} + \delta m_u,
        \overline{m} + \delta m_d,
        \overline{m} + \delta m_s )
     = X_S(\overline{m}, \overline{m},\overline{m})
               + O( (\delta m_q)^2 ) \,.
\label{XSconst}
\end{eqnarray}
There are many possibilities for singlet quantities. Using hadronic
masses we have, for example
\begin{eqnarray}
   X_N^2    &=& \sixth( M_p^2 + M_n^2 + M_{\Sigma^+}^2 +  M_{\Sigma^-}^2
                                 + M_{\Xi^0}^2 + M_{\Xi^-}^2 ) 
                    = (1.1610\,\mbox{GeV})^2
                                                              \nonumber   \\
   X_{\pi}^2 &=& \sixth( M_{K^+}^2 + M_{K^0}^2 + M_{\pi^+}^2 
                                + M_{\pi^-}^2 + M_{\overline{K}^0}^2 + M_{K^-}^2)
                    = (0.4116\,\mbox{GeV})^2
                                                              \nonumber   \\
   X_\rho^2 &=& \sixth(M_{K^{*+}}^2 + M_{K^{*0}}^2 + M_{\rho^+}^2 
                                + M_{\rho^-}^2 + M_{\overline{K}^{*0}}^2 + M_{K^{*-}}^2)
                             = (0.8562\,\mbox{GeV})^2 \,,
\end{eqnarray}
for octet baryons, pseudoscalar octet mesons and vector octet mesons
respectively.
Another baryon octet possibilty is $X_\Lambda^2 = \half(M_\Sigma^2 + M_\Lambda^2)$
but other singlet quantities can be constructed using the baryon decuplet.
Alternatively gluonic quantities can be used such as the
`Force' scale $X_{r_0}^2 = 1 / r_0^2$ or the Wilson flow scales,
introduced by L\"uscher
\begin{eqnarray}
   X_{t_0}^2 = {1 \over t_0}\,, \qquad  X_{w_0}^2 = {1 \over w_0^2}\,,
\end{eqnarray}
(see e.g.\ \cite{luscher10a,borsanyi12a}). These are all
`secondary scales', their physical value has to be determined.

The stationary point of $X_S$ can be checked, using the
Gell-Mann--Okubo $SU(3)$ flavour breaking expansion.
For example for the pseudoscalar octet mesons we have the expansion
\begin{eqnarray}
   M_{\pi^+}^2 (= M_{\pi^-}^2) &=& M_{0\pi}^2 + \alpha_\pi(\delta m_u + \delta m_d)
                           + O((\delta m_q)^2)
                                                    \nonumber \\
   M_{K^+}^2 (= M_{K^-}^2)    &=& M_{0\pi}^2 + \alpha_\pi(\delta m_u + \delta m_s)
                           + O((\delta m_q)^2)
                                                    \nonumber \\
   M_{K^0}^2(= M_{\overline{K}^0}^2) 
                           &=& M_{0\pi}^2 + \alpha_\pi(\delta m_d + \delta m_s)
                               + O((\delta m_q)^2) \,.
\end{eqnarray}
Constructing $X_\pi^2$ gives immediately the result of eq.~(\ref{stationary}).
Another check is to use $\chi$-PT (assuming that it is valid in the
neighbourhood of the $SU(3)$ flavour plane/line). Simply choose
your favourite $\chi$-PT result and expand about a $SU(3)$ flavour
symmetric line/point. For example in \cite{bar13a}, the chiral expansion
for $t_0$ (for mass degenerate $u$ and $d$ quarks) can be manipulated
\cite{bornyakov15a} to give
\begin{eqnarray}
   t_0 = T(\overline{\chi}) \left[ 
           1 + {1 \over (4 \pi f_0)^4}
               ( \textstyle{5 \over 6}k_2 
             + \textstyle{1 \over 4} k_5^{\prime\prime} ) (\chi_s - \chi_l)^2
                    + \cdots \right] \,,
 \end{eqnarray}
where $T$ is a (known) function of
$\overline{\chi} \equiv 1/3\,(2\chi_l + \chi_s)$ only.
As $(\chi_s - \chi_l) \propto (\delta m_s - \delta m_l)$ then this
agrees with our previous assertion: there is no linear term, the
first term is quadratic in $SU(3)$ flavour symmetry breaking.


\section{Lattice matters}


We have generated $2+1$ flavour gauge configurations using an action
consisting of tree level Symanzik glue and a mildy stout smeared
$O(a)$ non-perturbatively improved clover action, \cite{cundy09a},
at four-$\beta$ values, $\beta = 5.40, 5.50, 5.65, 5.80$ on a
variety of lattice sizes $24^3\times 48$, $32^3\times 64$ and $48^3\times 96$.
All box sizes have $L \gsim 2\,\mbox{fm}$. All the pion masses used
have $M_\pi L > 4$ and range from about $500$ to $220\,\mbox{MeV}$.
They are either at points on the $SU(3)$ flavour symmetric line
or along lines of constant $\overline{m}$. This gives $21$ data sets
at our disposal.

The quark mass $m_q$ and $\delta m_q$ are given by
\begin{eqnarray}
   m_q = {1 \over 2}
            \left( {1\over \kappa_q} - {1\over \kappa_{0c}} \right) \,, \qquad
   \delta m_q = m_q - \overline{m}
      = {1 \over 2}\left( {1 \over \kappa_q} 
                           - {1 \over \kappa_0} \right) \,,
\end{eqnarray}
where $\kappa_{0c}$ is chiral limit along symmetric line. (Note that
this cancels in $\delta m_q$.)

\begin{figure}[h]
\begin{center}

\begin{minipage}{0.45\textwidth}

   \begin{center}
      \includegraphics[width=7.00cm]
         {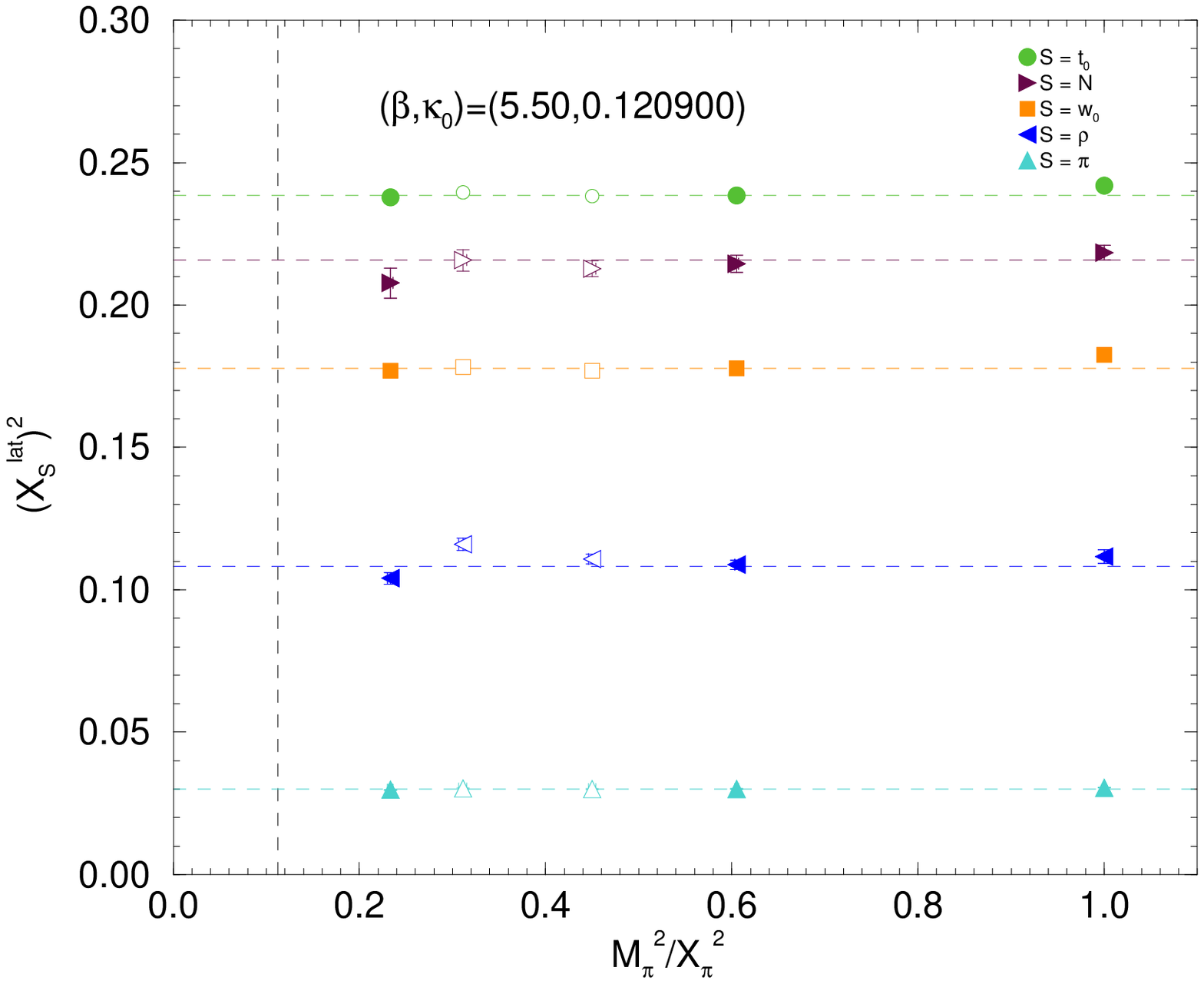}
   \end{center} 

\end{minipage}\hspace*{0.05\textwidth}
\begin{minipage}{0.45\textwidth}

   \begin{center}
      \includegraphics[width=7.00cm]
         {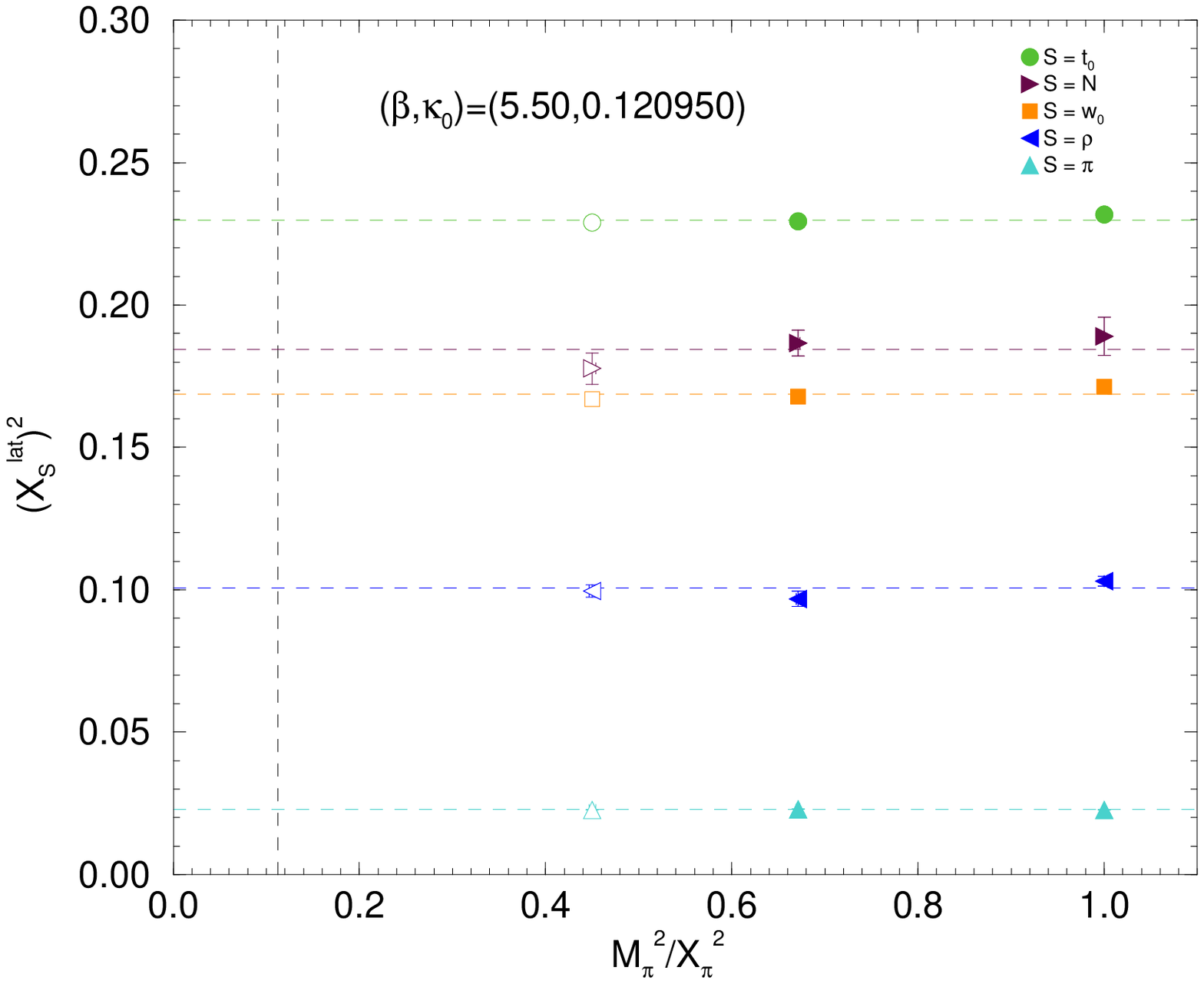}
   \end{center} 

\end{minipage}
\caption{Top to bottom $(X_S^{\rm lat})^2$ for $S = t_0$ (circles),
         $N$ (right triangles), $w_0$ (squares), $\rho$ (left triangles)
         and $\pi$ (up triangles) for $(\beta, \kappa_0) = (5.50, 0.120900)$
         (left panel) and $(\beta, \kappa_0) = (5.50, 0.120950)$ 
         (right panel) together with constant fits. The opaque
         points have $M_\pi L < 4$ and are not included in the fits.
         The vertical line represents the physical point.}
\label{XSres}

\end{center}
\end{figure}

We first investigate the constancy of singlet quantities, as
given in eq.~(\ref{XSconst}). In Fig.~\ref{XSres}
we plot $(X_S^{\rm lat})^2$ for $S = t_0$, $N$, $w_0$, $\rho$ and $\pi$
for $(\beta, \kappa_0) = (5.50, 0.120900)$, $(5.50, 0.120950)$. As
expected, in agreement with the discussion of section~\ref{singlet},
the $X_S^2$ singlet quantities are constant.

\begin{figure}[p]
\begin{center}

\begin{minipage}{0.45\textwidth}

   \begin{center}
      \includegraphics[width=7.00cm]
         {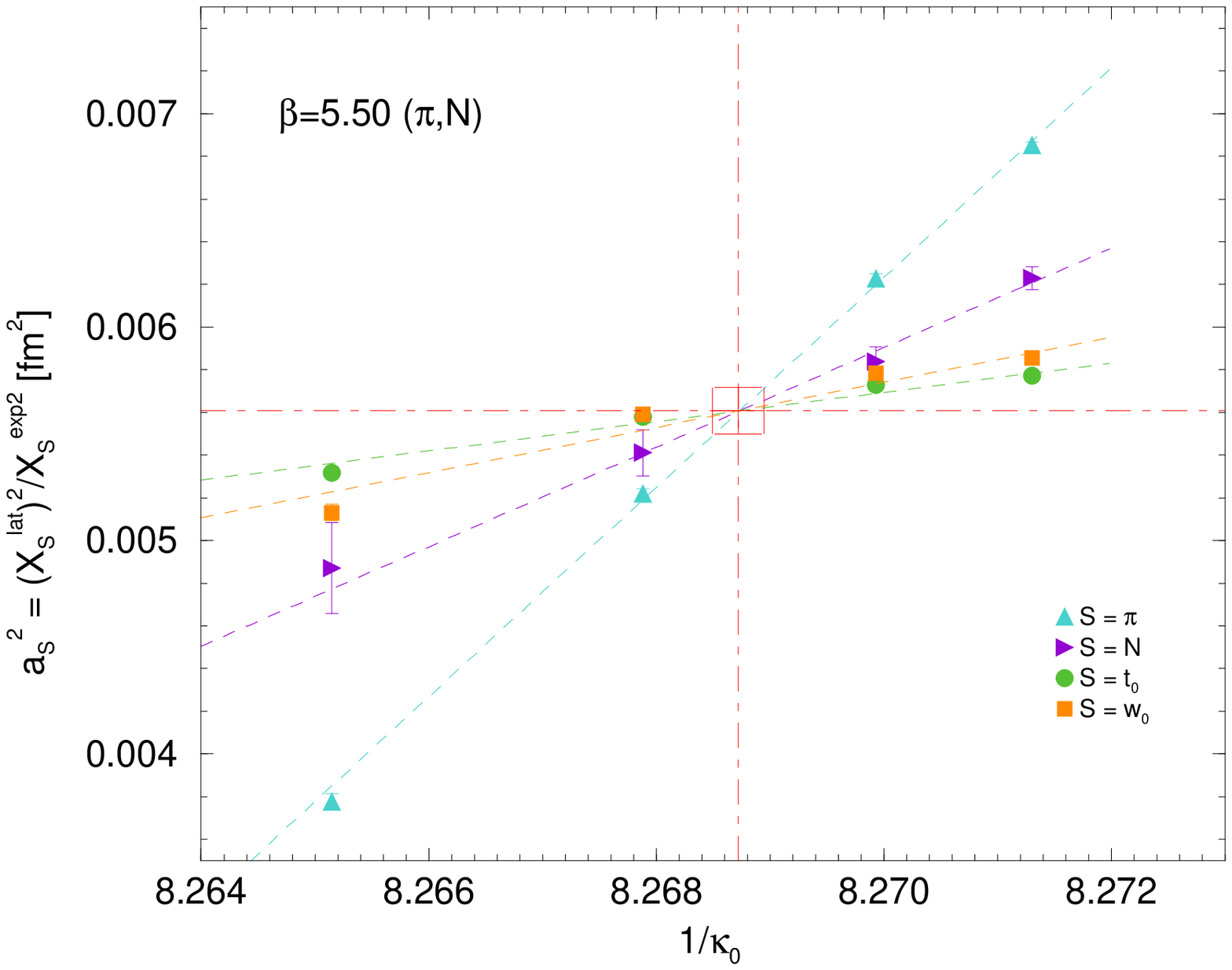}
   \end{center} 

\end{minipage}\hspace*{0.05\textwidth}
\begin{minipage}{0.45\textwidth}

   \begin{center}
      \includegraphics[width=7.00cm]
         {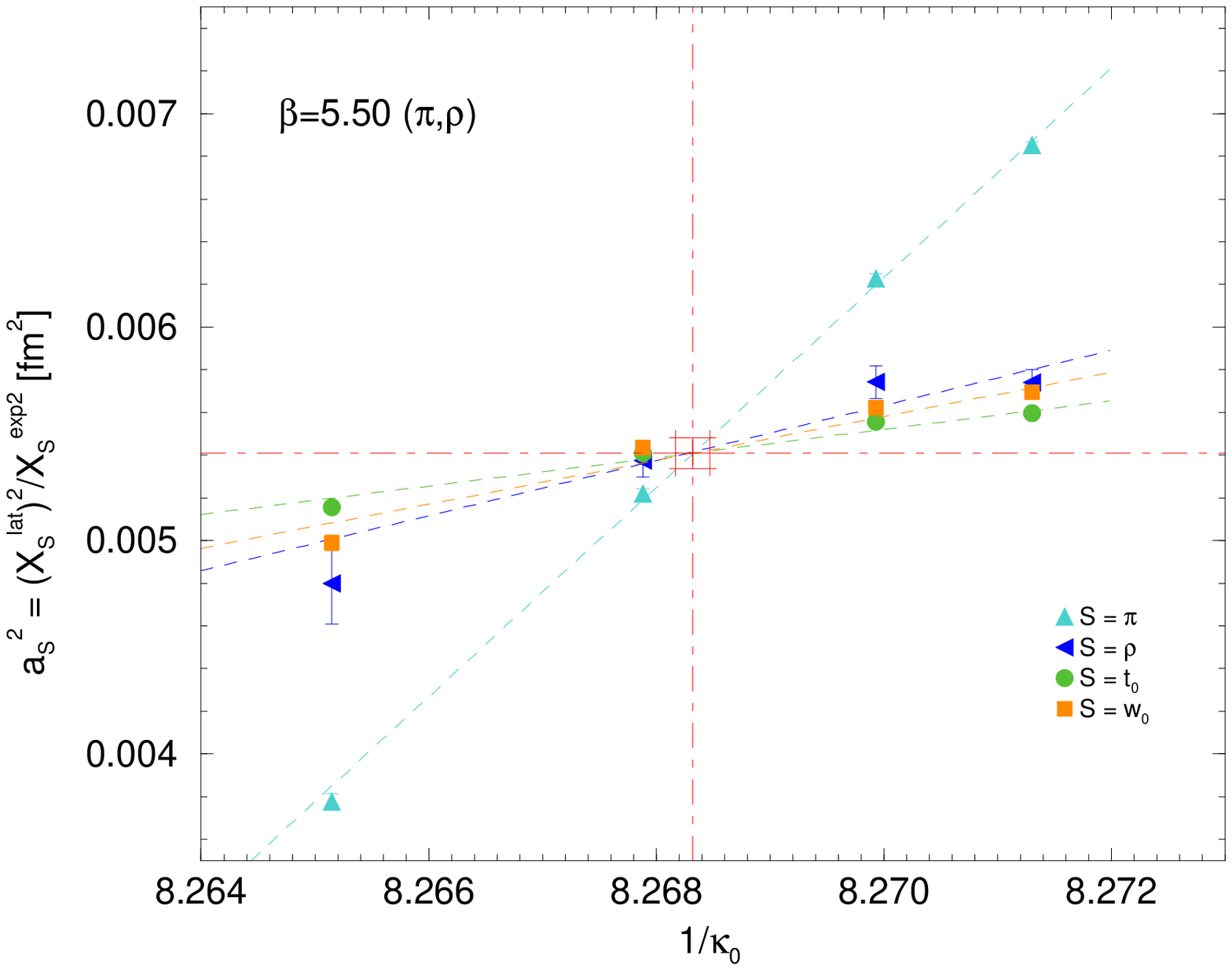}
   \end{center} 

\end{minipage}
\caption{$a_S^2$ against $1/\kappa_0$ for $S = \pi$, $N$ and $t_0$, $w_0$ 
         together with quadratic fits for $\beta = 5.50$. 
         Left panel: based on $(\pi, N)$ crossing; 
         Right panel: based on $(\pi, \rho)$ crossing.}
\label{crossing}

\end{center}
\end{figure}

\begin{figure}[p]
\begin{center}

\begin{minipage}{0.45\textwidth}

   \begin{center}
      \includegraphics[width=7.00cm]
         {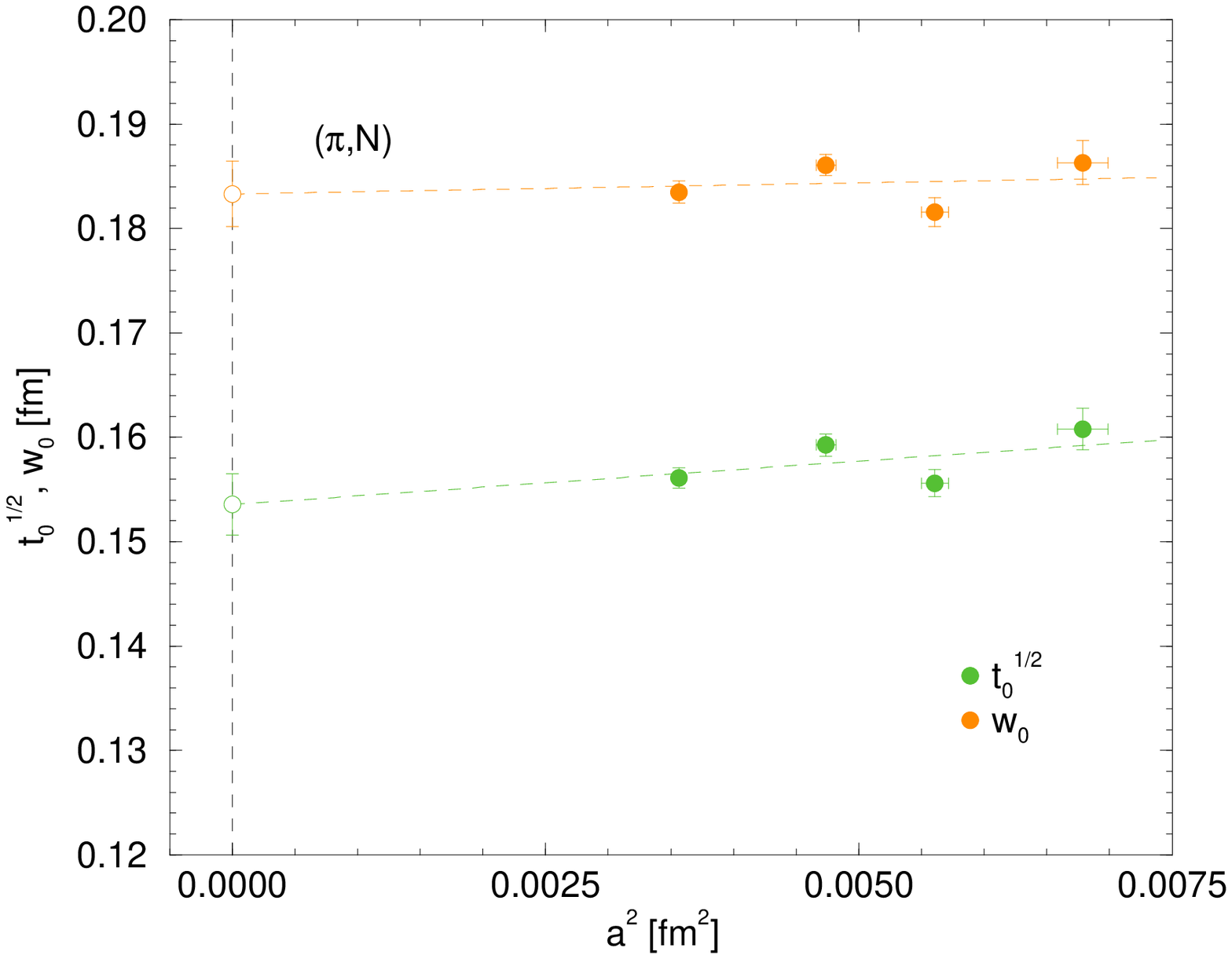}
   \end{center} 

\end{minipage}\hspace*{0.05\textwidth}
\begin{minipage}{0.45\textwidth}

   \begin{center}
      \includegraphics[width=7.00cm]
         {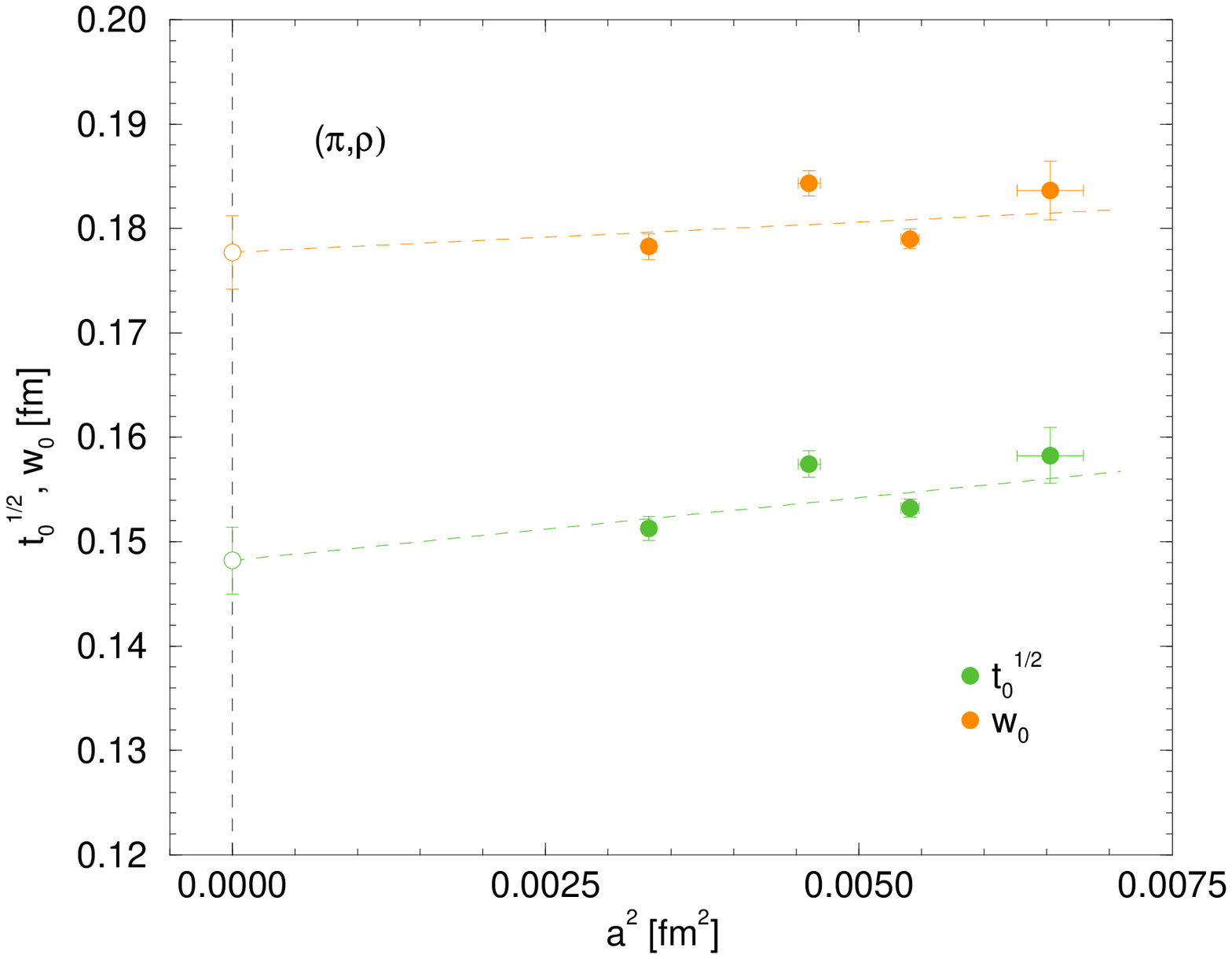}
   \end{center} 

\end{minipage}
\caption{$\sqrt{t_0}$ and $w_0$ (in $\mbox{fm}$) against $a^2$
         (in $\mbox{fm}^2$) from the $(\pi, N)$ crossing (left panel)
         and $(\pi, \rho)$ (right panel) crossing together
         with a linear fit.}
\label{continuum}

\end{center}
\end{figure}

We now take $X_S = \mbox{const.}$ to determine the scale
\begin{eqnarray}
   a_S^2(\kappa_0) = { X_S^{\rm lat\,2}(\kappa_0) \over X_S^{\exp\,2} } \,.
\label{as2}
\end{eqnarray}
This is a function of $m_0$ or here $\kappa_0$. So if we vary $\kappa_0$
(for example as in Fig.~\ref{XSres}) -- when pairs $a_S$, $a_{S^\prime}$
cross this gives a common lattice spacing $a$. We apply this in
particular here to%
\footnote{For the $\beta$ and pion mass values considered here,
the $\rho$ and $K^*$ are stable particles.}
\begin{eqnarray}
   (S, S^\prime) = (\pi,N), \, (\pi,\rho) \,.
\end{eqnarray}
For $S = t_0$, $w_0$ we can arrange $X_{t_0}^{\exp}$, $X_{w_0}^{\exp}$
(from eq.~(\ref{as2})) so that these singlet quantities also cross
at the same point. In Fig.~\ref{crossing} we show these crossings
for $\beta = 5.50$. From the results for the four beta values we
can now make the last, continuum extrapolation. This is shown in
Fig.~\ref{continuum}.
A (weighted) average of these results gives our final estimates
for $\sqrt{t_0^{\exp}}$, $w_0^{\exp}$ as found in \cite{bornyakov15a}.

\begin{figure}[p]
\begin{center}

\begin{minipage}{0.45\textwidth}

   \begin{center}
      \includegraphics[width=7.00cm]
         {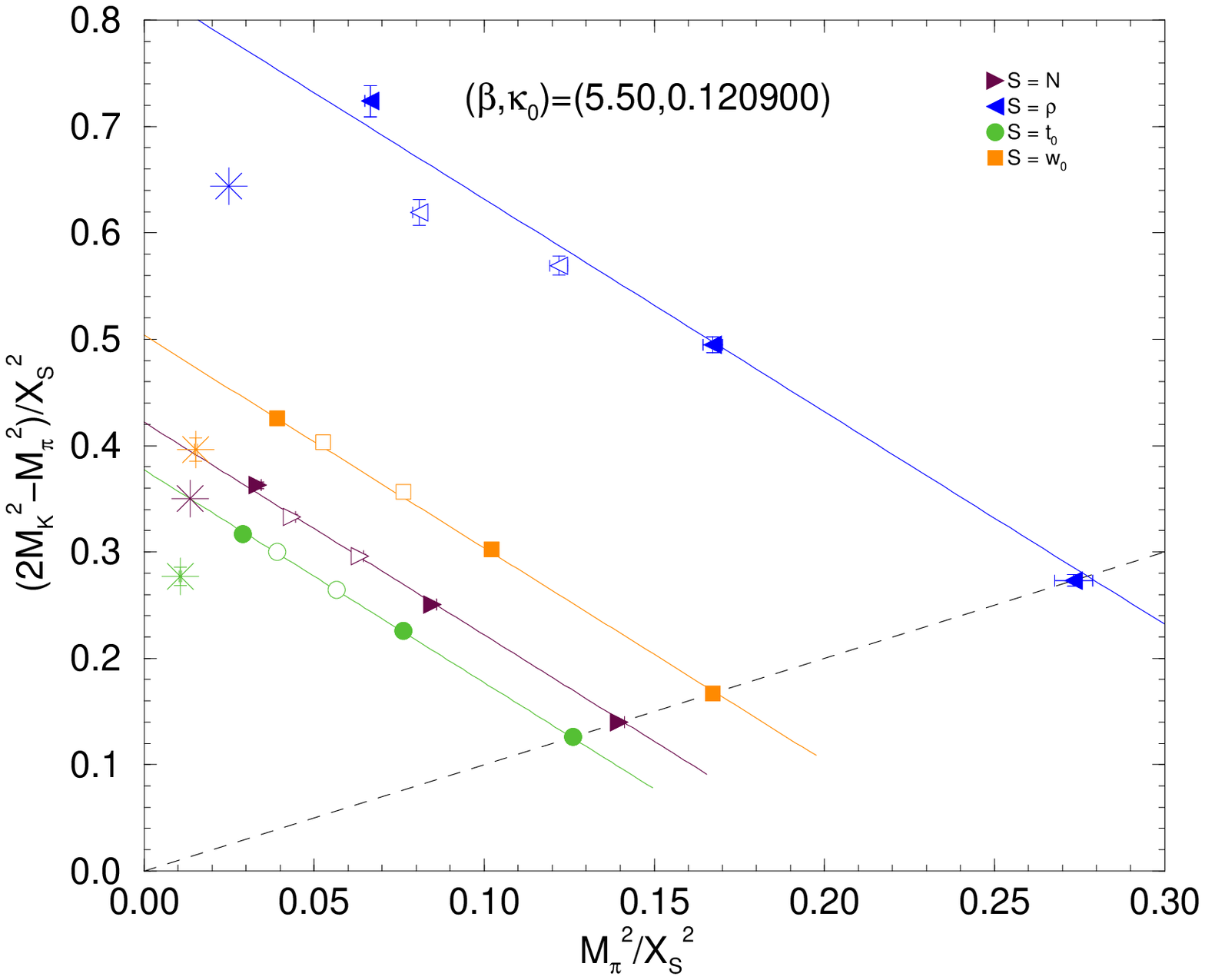}
   \end{center} 

\end{minipage}\hspace*{0.05\textwidth}
\begin{minipage}{0.45\textwidth}

   \begin{center}
      \includegraphics[width=7.00cm]
         {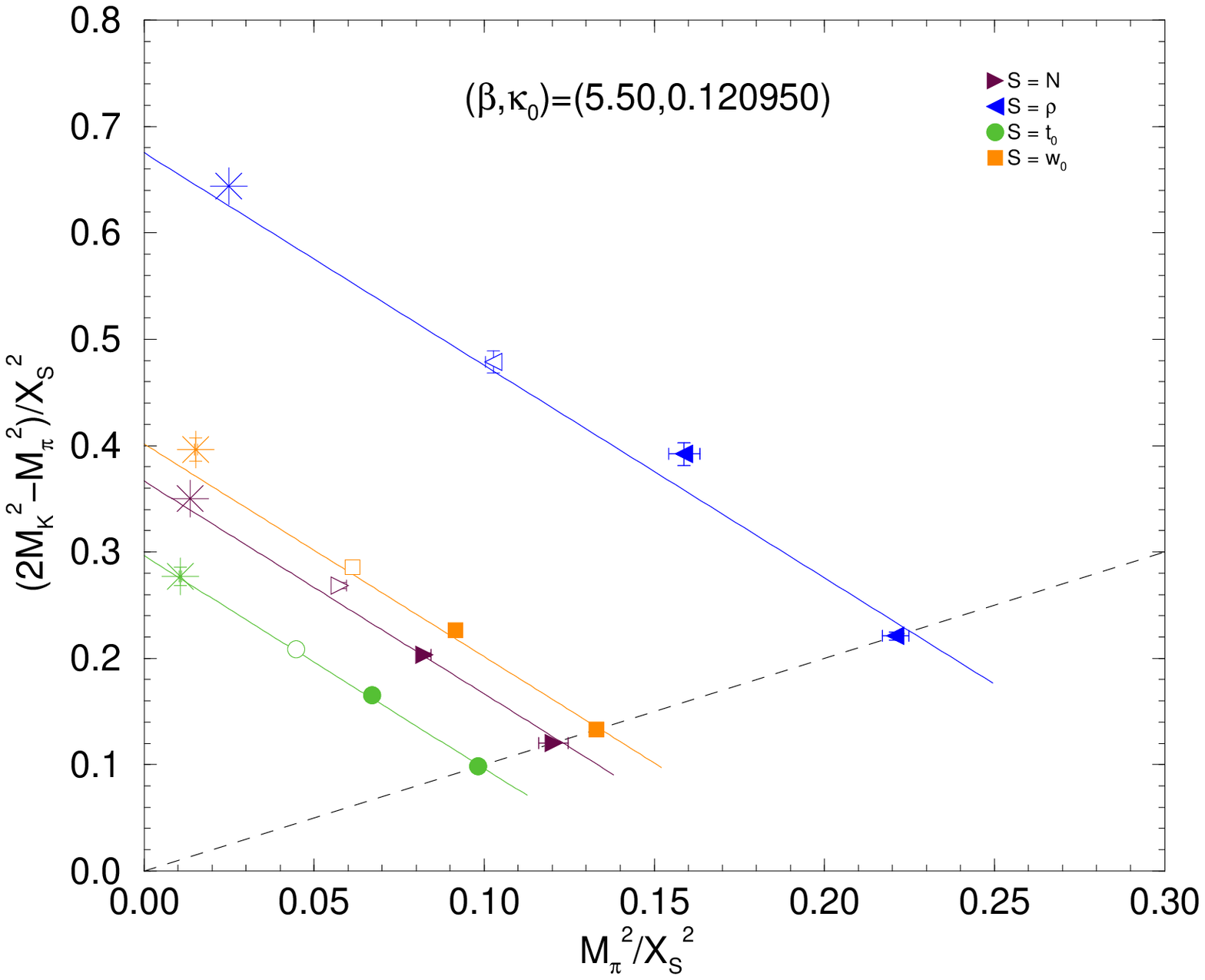}
   \end{center} 

\end{minipage}
\caption{$(2M_K^2 - M_\pi^2) / X^2_S$ against $M_\pi^2 / X_S^2$,
         together with the fit from eq.~(\protect\ref{qm_plane})
         for $(\beta, \kappa_0) = (5.50, 0.120900)$ (left panel)
         and $(5.50, 0.120950)$ (right panel). The stars correspond
         to the phenomenological values.}
\label{derived_path}

\end{center}
\end{figure}

Alternatively we can write
\begin{eqnarray}
   {2M_K^2 - M_\pi^2 \over X^2_S} 
       = C - 2{M_\pi^2 \over X_S^2} \,,
\label{qm_plane}
\end{eqnarray}
($C = X_\pi^2 / X_S^2$) for $S = N, \rho, t_0, w_0$.
In Fig.~\ref{derived_path} we
plot this function for $(\beta, \kappa_0) = (5.50, 0.120900)$,
$(5.50, 0.120950)$. This represents the path in the quark mass plane.
Also shown are the experimental values (now including those of
$S = t_0$, $w_0$). We see that these $\kappa_0$ values straddle 
the optimum $\kappa_0^*$ -- it is clear that $\kappa_0^*$ lies closer
to $0.120950$ than $0.120900$.

\begin{figure}[h]
\begin{center}

\begin{minipage}{0.45\textwidth}

   \begin{center}
      \includegraphics[width=7.00cm]
         {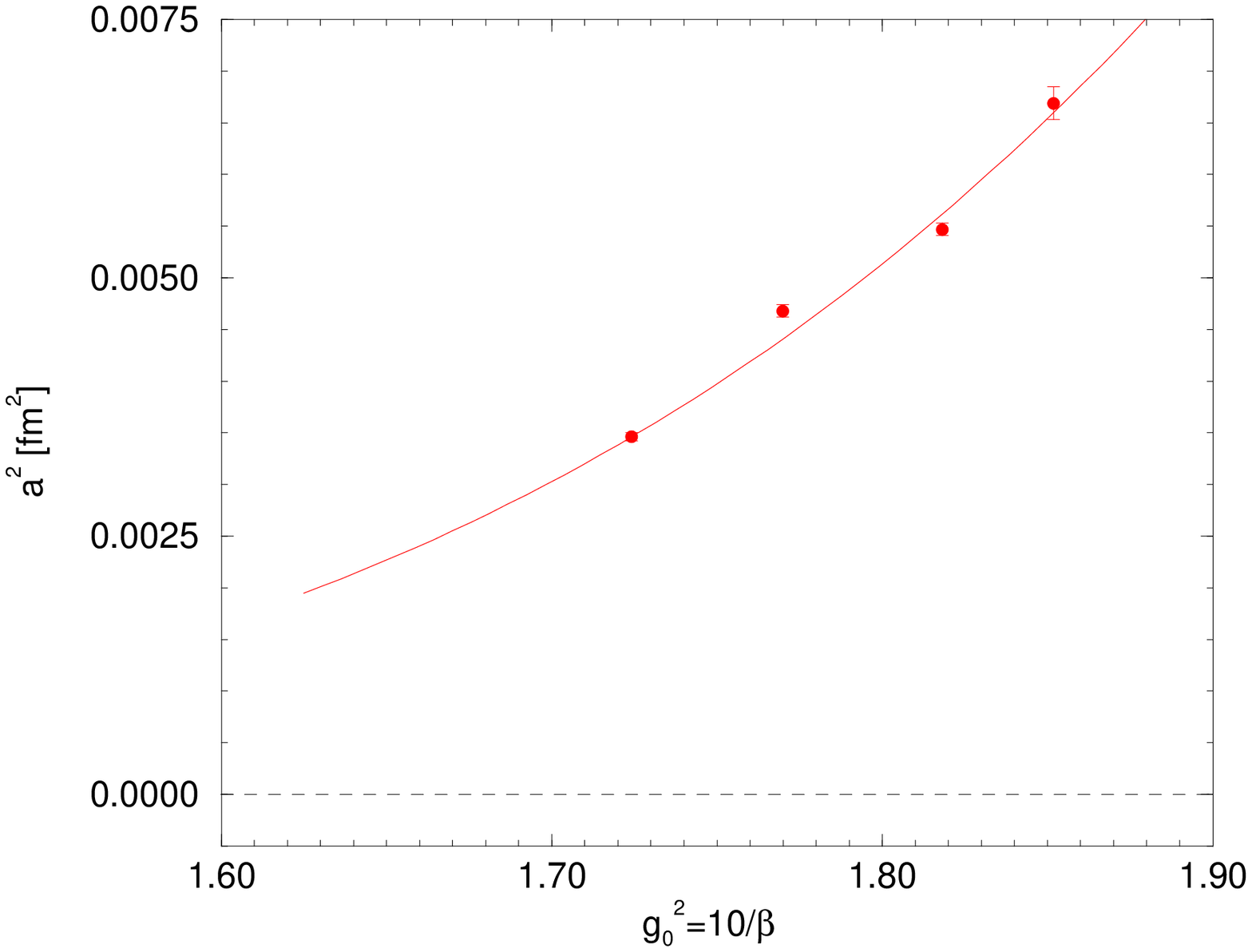}
   \end{center} 

\end{minipage}\hspace*{0.05\textwidth}
\begin{minipage}{0.45\textwidth}

   \begin{center}
      \includegraphics[width=7.00cm]
         {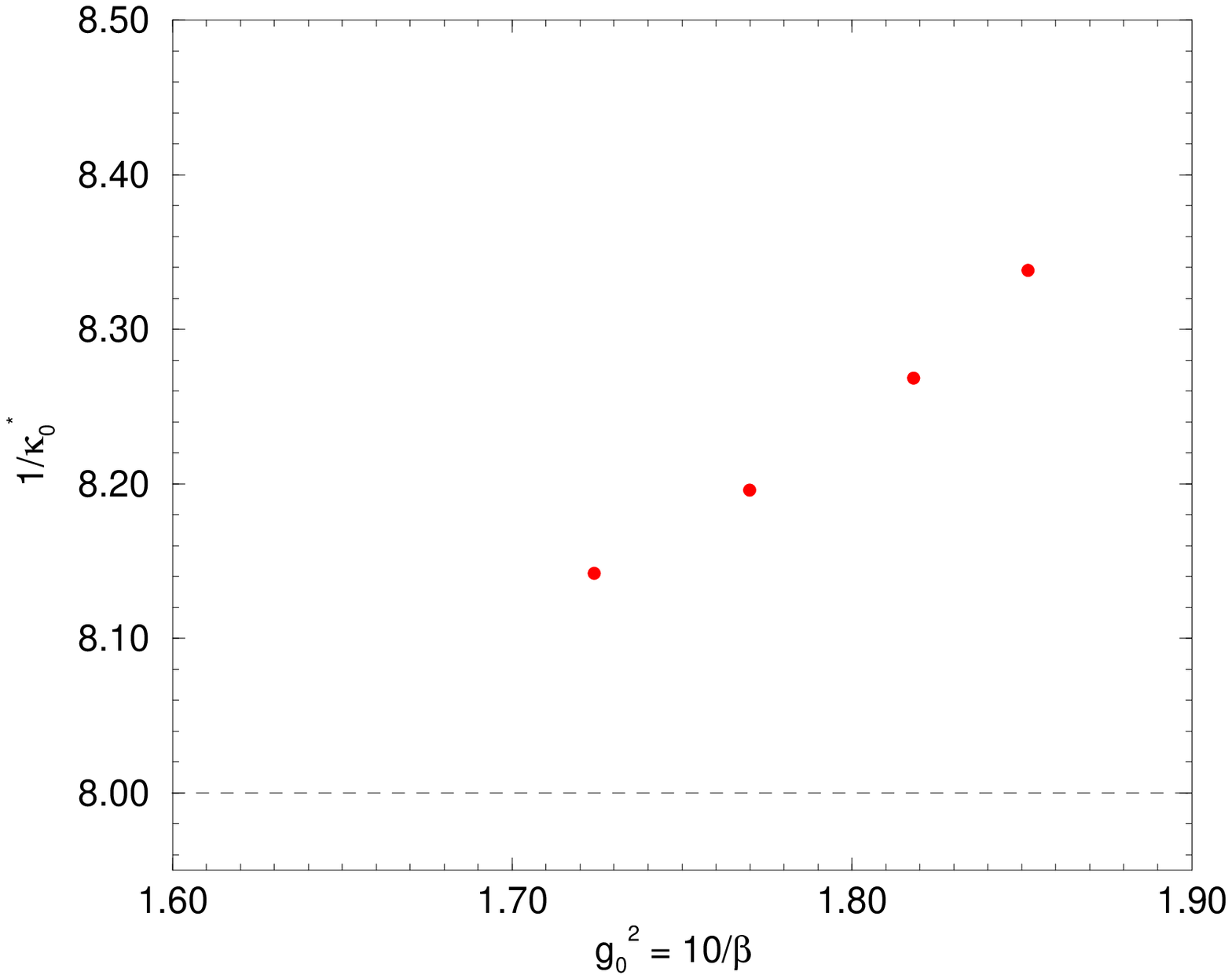}
   \end{center} 

\end{minipage}
\caption{Left panel: $a^2$ versus $g_0^2$. The curve is from the
         running coupling constant, using the $2$-loop beta function
         normalised to the $\beta = 5.80$ result.
         Right panel: $1/\kappa_0^*$ versus $g_0^2$. The horizontal
         dashed lines represents the value in the continuum limit.}
\label{scaling+comparison}

\end{center}
\end{figure}

Finally we comment on our results. In the left panel of
Fig.~\ref{scaling+comparison} we plot $a^2$ against $g_0^2$.
The curve is the running coupling constant, $g_0^2 = 10/\beta$
using the $2$-loop QCD beta function, normalised to $\beta = 5.80$, namely
\begin{eqnarray}
   {a^2(\beta) \over a^2(\beta_0)}
      = \left({\beta_0 \over \beta}\right)^{-{b_1 \over b_0^2}} \,
        \exp\left( {-{1\over 10b_0}\, (\beta - \beta_0)}\right) \,,
   \qquad \beta_0 = 5.80 \,,
\end{eqnarray}
($b_0$, $b_1$ are the first two coefficients of the beta function).
There seems to be reasonable agreement between the data points and
the curve. The right hand panel of Fig.~\ref{scaling+comparison}
indicates how the initial point, $\kappa_0^*$, on the $SU(3)$ flavour
symmetric line changes with $g_0^2$.


\section{Conclusions}

\begin{figure}[h]
\begin{center}
   \includegraphics[width=6.00cm]{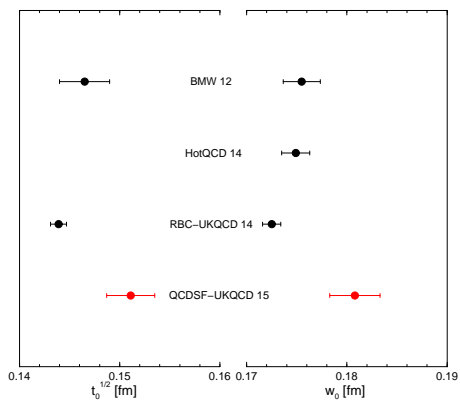}
\end{center} 
\caption{$\sqrt{t_0^{\exp}}$, left plot and $w_0^{\exp}$,
         right plot in $\mbox{fm}$ for BMW 12 \protect\cite{borsanyi12a},
         HotQCD 14 \protect\cite{bazavov14a}, 
         RBC-UKQCD 14 \protect\cite{blum14a}, together with the
         present results.}
\label{comparison}
\end{figure}

Our programme is to tune strange and light quark masses to their physical
values simultaneously by keeping
$\overline{m} = 1/3 \, ( 2m_l + m_s ) = \mbox{const.}$.
As the light quark mass is decreased then $M_\pi \searrow$ and $M_K \nearrow$.
Singlet quantities, here denoted by $X_S(\kappa_0)$ remain constant
starting from a point on the $SU(3)$ flavour symmetric 
line --- the Gell-Mann--Okubo result. We can use this result and
$X_S^{\exp}$ to determine the $a_S(\kappa_0)$ scale. Varying $\kappa_0$ 
-- determines when pairs of singlet quantities such as $(X_\pi,X_N)$
and $(X_\pi,X_\rho)$ cross giving a common lattice spacing $a$.
By arranging so that $X_{t_0}$, $X_{w_0}$ also cross here,
we are able to give a determination of the `secondary' scales
$\sqrt{t_0^{\exp}}$ and $w_0^{\exp}$ $[\mbox{fm}]$. Finally in
Fig.~\ref{comparison} a comparison  with other results is given.


\section*{Acknowledgements}


The numerical configuration generation 
(using the BQCD lattice QCD program \cite{nakamura10a})
and data analysis 
(using the Chroma software library \cite{edwards04a})
was carried out
on the IBM BlueGene/Q using DIRAC 2 resources (EPCC, Edinburgh, UK),
the BlueGene/P and Q at NIC (J\"ulich, Germany),
the Lomonosov at MSU (Moscow, Russia) and the
SGI ICE 8200 and Cray XC30 at HLRN (The North-German Supercomputer
Alliance) and on the NCI National Facility in Canberra, Australia
(supported by the Australian Commonwealth Government).
HP was supported by DFG Grant No. SCHI 422/10-1.
PELR was supported in part by the STFC under contract ST/G00062X/1
and JMZ was supported by the Australian Research Council Grant
No. FT100100005 and DP140103067. We thank all funding agencies.



\end{document}